\documentclass[10pt,letterpaper]{article}
\usepackage[top=0.85in,left=2.75in,footskip=0.75in]{geometry}
\usepackage[utf8]{inputenc}
\usepackage[T1]{fontenc}
\usepackage{amsmath,amssymb}
\usepackage{changepage}
\usepackage{textcomp,marvosym}
\usepackage{cite}
\usepackage{nameref,hyperref}
\hypersetup{hidelinks}
\usepackage[nopatch=eqnum]{microtype}
\DisableLigatures[f]{encoding = *, family = * }
\usepackage[table]{xcolor}
\usepackage{array}
\usepackage{booktabs}
\usepackage{multirow}
\raggedright
\usepackage{placeins}

\usepackage[aboveskip=1pt,labelfont=bf,labelsep=period,justification=raggedright,singlelinecheck=off]{caption}

\makeatletter
\renewcommand{\@biblabel}[1]{\quad#1.}
\makeatother
\usepackage{lastpage,fancyhdr,graphicx}
\usepackage{float}
\fancyheadoffset[L]{2.25in}
\fancyfootoffset[L]{2.25in}
\begin{document}
\vspace*{0.2in}

\begin{flushleft}
{\Large
\textbf\newline{What makes vertical federated learning work with thirteen samples? A controlled benchmark on coral thermal-stress multi-omics}
}
\newline
\\
Sam Victor\textsuperscript{*}
\\
\bigskip
Independent Researcher
\\
\bigskip
* samvictordr@outlook.com
\end{flushleft}

\section*{Abstract}
Vertical federated learning (VFL) lets laboratories that hold different omics layers of the same samples train a joint model without sharing raw data, but it is usually developed for datasets with thousands of samples.
We benchmarked VFL on a \textit{Montipora capitata} thermal-stress dataset with 13 samples and 90,579 features across transcriptomics, proteomics, metabolomics and microbiome data.
The benchmark crosses two VFL backbones (a split-network model implemented in NVIDIA FLARE, and LASER-VFL) with three per-silo feature-selection strategies (gradient saliency, random, univariate $t$-test) and two allocations of a 1,300-feature budget, and adds a centralised elastic-net reference.
All models were trained for the same number of steps and evaluated by repeated stratified 3-fold and leave-one-out cross-validation, with separate estimates of seed and data-split variability and end-to-end label-permutation tests.
Plain VFL on all features classified heat stress well above chance (3-fold AUROC 0.869, 95\% CI 0.838--0.895; permutation $p = 0.020$).
Gradient-saliency selection with hand-set omics priors (0.904) could not be distinguished from plain VFL, from equal budgets or from random subsets of the same size (corrected $p \geq 0.56$), and its selected features were no more consistent across seeds than random ones.
LASER-VFL failed on all features at a learning rate of $10^{-3}$ but matched the other methods at $10^{-4}$.
This version corrects versions 1--3 of this preprint, which reported that standard VFL performed at chance and that domain priors stabilised training; we show that those results arose because every model selected its checkpoint on a single-sample validation set, which evaluated the baseline after one optimiser step.
With very few samples, evaluation choices outweighed every method choice we tested, and we give recommendations for evaluating federated models in this regime.

\section*{Author summary}
Coral reefs are threatened by marine heatwaves, and understanding how corals respond requires combining several kinds of molecular measurements that are often produced by different laboratories.
Federated learning could let those laboratories build a shared model without pooling their raw data, but coral studies usually have very few samples.
We tested how well federated learning works on a coral heat-stress dataset with only 13 samples and about 90,000 measurements per sample.
A standard federated model already worked well; adding expert knowledge about which data types matter, or selecting measurements with an attribution method, did not help measurably, and selecting measurements at random worked just as well.
An earlier version of this study reached the opposite conclusion because of a subtle error in how the models were evaluated, which we explain and correct.
Our results suggest that, when data are this scarce, careful and equal evaluation of every method matters more than designing new methods.

\section*{Introduction}

Mass coral bleaching driven by marine heatwaves has become a recurrent, global phenomenon \cite{hughes2017global,hughes2018spatial}.
Understanding why some colonies tolerate heat stress better than others requires measuring the coral holobiont---the coral host, its algal symbionts and its bacterial community---at several molecular levels at once \cite{voolstra2021extending,ziegler2017bacterial,barshis2013genomic,dixon2015genomic}.
Multi-omic profiling of thermally stressed corals is therefore increasingly common \cite{cziesielski2018multiomics,williams2021multiomic}, and machine-learning models have been trained on such data to predict stress status \cite{mayfield2022proteomic}.
In practice the different omics layers of a study are often generated by different laboratories or core facilities, and integrating them computationally is a standing challenge for multi-omics analysis \cite{hasin2017multiomics,subramanian2020multiomics,picard2021integration}.

Integration methods such as DIABLO \cite{singh2019diablo,rohart2017mixomics}, MOFA \cite{argelaguet2018mofa} and MOGONET \cite{wang2021mogonet} assume that all omics layers are pooled in one place.
When layers are held by different parties that cannot share raw data, vertical federated learning (VFL) offers an alternative: each party keeps its feature block and trains a local encoder, and only intermediate representations and their gradients are exchanged with a party that holds the labels \cite{yang2019fml,liu2024vfl,vepakomma2018split,gupta2018distributed}.
Federated learning has been demonstrated in medicine \cite{rieke2020future,sheller2020federated,dayan2021exam,warnat2021swarm} and for omics data \cite{zhou2024ppmlomics}, and recent VFL methods address communication cost \cite{castiglia2022cvfl}, feature selection \cite{castiglia2023lessvfl} and missing feature blocks \cite{valdeira2025laser}.
These methods are typically developed and evaluated on datasets with thousands of samples.

Coral multi-omics studies sit at the opposite extreme.
The dataset studied here has 13 samples with complete data in all four omics layers and 90,579 features, a ratio of features to samples of nearly 7,000.
In this regime almost every evaluation choice matters.
Error bars of cross-validated estimates are wide \cite{varoquaux2018cv,vabalas2019validation}; feature selection or model selection performed outside the cross-validation loop inflates performance \cite{ambroise2002selection,simon2003pitfalls,varma2006bias,cawley2010overfitting,kapoor2023leakage}; leave-one-out estimates of the area under the ROC curve (AUROC) are biased \cite{parker2007stratification,airola2011auc}; and the variance due to random initialisation can be as large as the differences between methods \cite{bouthillier2021variance,henderson2018deep}.
Neural baselines that are not tuned or trained with the same care as a proposed method can make the proposed method look better than it is \cite{melis2018state,lipton2019troubling}.
Neutral benchmarks that control these factors are needed before any method can be recommended \cite{boulesteix2013plea,weber2019essential,boulesteix2017towards}.

\paragraph{Correction of an earlier version of this work.}
Versions 1--3 of this preprint proposed REEF (hereafter, the saliency pipeline), a VFL pipeline that combined gradient-saliency feature selection with hand-set ``biological'' priors over omics layers, and reported that it stabilised VFL where a standard VFL baseline performed at chance (AUROC 0.500) and LASER-VFL \cite{valdeira2025laser} was unstable.
Re-running that code on new hardware, we traced the baseline failure to an evaluation bug: every method selected its checkpoint on a validation set of a single sample, for which AUROC is undefined and was returned as a constant 0.5.
As a result the standard VFL baseline was always evaluated at its first training round, the saliency pipeline at its 21st, and LASER---which stopped on validation loss instead---at a variable epoch.
The conclusions of versions 1--3 do not survive correcting this bug (see \nameref{sec:reproduction}).
This version replaces them with a controlled benchmark designed to answer the question the earlier version asked: what, if anything, makes VFL work with this few samples?

\paragraph{Contributions.}
\begin{enumerate}
\item A factorial benchmark that crosses two VFL backbones (a split-network VFL model implemented in NVIDIA FLARE \cite{roth2022nvflare} and LASER-VFL \cite{valdeira2025laser}) with three per-party feature-selection strategies (gradient saliency, random, univariate $t$-test) and two feature-budget allocations (expert prior, equal), with a centralised elastic-net reference \cite{zou2005elastic}.
\item An evaluation protocol for $N \approx 10$ that fixes training length, matches optimiser steps across methods, separates seed variance from data-split variance, uses label-permutation tests, and measures the stability of selected features \cite{nogueira2018stability}.
\item A root-cause analysis and correction of the results in versions 1--3, including a reproduction of the original numbers with the original code.
\end{enumerate}

\section*{Materials and methods}

\subsection*{Dataset}

We used the \textit{Montipora capitata} thermal-stress multi-omics dataset distributed as a single workbook on Zenodo \cite{stephens2022zenodo}.
The data come from a five-week tank experiment in which nubbins of four colonies from K\=ane\textquoteleft ohe Bay, Hawai\textquoteleft i, were kept at ambient temperature (about 27.5\,\textdegree C) or heated to about 3\,\textdegree C above ambient, and sampled at five time points; nubbins were also collected from the source colonies in the field \cite{williams2021metabolomic,williams2021multiomic,williams2023peeling}.
Transcriptomes and metabolomes were first analysed in \cite{williams2021metabolomic,williams2021multiomic} and proteomes and 16S rRNA microbiomes in \cite{williams2023peeling}.
Time points T1, T3 and T5 correspond to the end of the temperature ramp, the onset of bleaching and the end of the experiment \cite{williams2021multiomic}.
Transcriptomes and proteomes were generated for colony MC-289 only, whereas metabolomes and microbiomes cover all four colonies \cite{williams2023peeling}.
Four omics layers were used, each treated as one party (silo) in the VFL setting: transcriptomics (62,038 predicted genes; Data S3), proteomics (4,054 proteins; Data S1; 123 missing values set to zero), metabolomics (12,055 features; Data S2) and the 16S rRNA microbiome table after the original authors' filtering (12,432 amplicon sequence variants; Data S5).
Only 13 samples, all from colony MC-289, are present in all four layers (Table~\ref{tab:samples}).
Williams et al.\ reported that two of these transcriptomes (T5-HiT\_2998 and T5-Amb\_1721) share fewer than 90\% of single-nucleotide polymorphisms with the other MC-289 samples, possibly because of a labelling mix-up \cite{williams2023peeling}; we therefore repeated the main analysis without them.
The binary label is high-temperature treatment (HiT; 6 samples) versus not (7 samples: 5 ambient-tank samples and 2 field samples).
Values were standardised per feature inside every training fold; no other transformation was applied.

\begin{table}[!ht]
\centering
\caption{{\bf The 13 samples present in all four omics layers.} All come from colony MC-289. T1, T3 and T5 are sampling time points of the tank experiment \cite{williams2021multiomic}.}
\label{tab:samples}
\begin{tabular}{lccc}
\toprule
Group & Treatment & $n$ & Label \\
\midrule
Field & field-collected & 2 & 0 \\
T1 & ambient / high temperature & 1 / 2 & 0 / 1 \\
T3 & ambient / high temperature & 2 / 2 & 0 / 1 \\
T5 & ambient / high temperature & 2 / 2 & 0 / 1 \\
\midrule
Total & & 13 & 7 / 6 \\
\bottomrule
\end{tabular}
\end{table}

\subsection*{Vertical federated learning setting}
Each silo $k \in \{1,\dots,4\}$ holds the feature block $x_k$ of every sample; the server holds the labels.
In each round the silos send embeddings $h_k = f_k(x_k)$ to the server, the server computes the loss of a fusion model $g([h_1;\dots;h_4])$ and returns $\partial L / \partial h_k$, and each silo updates its encoder \cite{vepakomma2018split,liu2024vfl}.
All methods were trained full-batch, because a training fold contains at most 12 samples.

\subsection*{Backbones}
\paragraph{Split-network VFL (``VFL'').}
The baseline of versions 1--3, implemented with the NVIDIA FLARE 2.7.1 data-exchange API \cite{roth2022nvflare}.
Each encoder is a multilayer perceptron $d_k \rightarrow d_\text{mid} \rightarrow 128 \rightarrow 64$ with $d_\text{mid} = \max(\min(d_k/4, 512), 128)$, layer normalisation \cite{ba2016layernorm}, ReLU and dropout 0.3 \cite{srivastava2014dropout}, initialised with Xavier-normal weights (gain 0.1) \cite{glorot2010understanding}.
The server concatenates the four 64-dimensional embeddings and applies a $256 \rightarrow 128 \rightarrow 64 \rightarrow 1$ classifier.
Encoders and fusion head use AdamW \cite{loshchilov2019adamw} (learning rates $5\times10^{-5}$ and $5\times10^{-4}$, weight decay $10^{-4}$) and client-side gradient clipping at norm 1.

\paragraph{LASER-VFL.}
LASER-VFL trains a family of predictors over subsets of parties by sharing parameters and sampling tasks, so that predictions remain possible when blocks are missing \cite{valdeira2025laser}.
We used the feature extractor, fusion model and task-sampling code of the authors' repository (commit \texttt{b7ff159}), with a hidden width of 128, 64-dimensional embeddings, AdamW (learning rate $10^{-3}$, weight decay $10^{-5}$) and the numerical safeguards described in S1 Text (layer normalisation of embeddings, small initialisation of classification heads, gradient clipping at norm 1).
No feature blocks were missing ($p_\text{miss}=0$); predictions use the first head's full-party output.
LASER-VFL was designed for missing features rather than for $P \gg N$, so it is included as a strong general-purpose VFL method, not as a method that is expected to suit this regime.

\subsection*{Feature selection inside the federation}
Each silo selects a subset of its own features inside every training fold; test samples never enter selection.
Two budget allocations give 1,300 selected features in total:
an \emph{expert} allocation used in versions 1--3 (transcriptomics 150, proteomics 250, metabolomics 500, microbiome 400), and an \emph{equal} allocation (325 per silo).
Three selectors were compared.
\begin{itemize}
\item \textbf{Gradient saliency (versions 1--3).} The VFL model is trained on all features for 20 rounds; each silo then scores feature $j$ by the mean over training samples of $\sum_{i=1}^{64} |\partial h_{k,i} / \partial x_{k,j}|$ and keeps its top-scoring features \cite{simonyan2014saliency,ancona2018towards}. Training continues for 30 rounds on the selected features.
\item \textbf{Random.} Each silo keeps a uniformly random subset of the budgeted size.
\item \textbf{Univariate $t$-test.} Each silo keeps the features with the largest absolute Welch $t$ statistic between classes in the training fold \cite{saeys2007review}. This requires the labels at the silo, which VFL normally keeps at the server, so it is included as a label-aware reference rather than as a deployable federated method.
\end{itemize}
For the VFL backbone, every selector uses the same 20 + 30 round schedule, so that the arms differ only in which features are kept.
The saliency pipeline of versions 1--3 additionally multiplies each silo's embedding by a fixed weight (transcriptomics 1.5, proteomics 1.0, metabolomics 0.8, microbiome 0.5); the ``equal'' arms use weight 1 for every silo.
For the LASER backbone, the selected subsets (from saliency selection in the same fold and seed, random, or $t$-test) are given to LASER-VFL, which is then trained from scratch.

\subsection*{Centralised reference}
An elastic-net logistic regression \cite{zou2005elastic,tibshirani1996lasso} ($\ell_1$ ratio 0.5, $C=1$, SAGA solver) on the concatenated, standardised features gives a non-federated reference that is deterministic given the data split \cite{pedregosa2011sklearn}.

\subsection*{Training protocol}
Every neural model was trained for a fixed 50 optimiser steps (50 full-batch rounds or epochs) and the final model was used for prediction.
No early stopping or checkpoint selection was performed, because a validation set carved out of 12 training samples cannot rank checkpoints reliably \cite{prechelt1998early,cawley2010overfitting}.
All arms within a fold share the same training samples and the same random seed.
No hyperparameters were tuned for any arm; the effect of training length and of LASER-VFL's learning rate is reported as a sensitivity analysis.

\subsection*{Evaluation}
\paragraph{Cross-validation schemes.}
(i) Leave-one-out cross-validation (LOOCV) with the 13 held-out predictions pooled into one AUROC per seed, repeated for 20 training seeds.
(ii) Stratified 3-fold cross-validation repeated over 10 random partitions, each run with 10 training seeds; AUROC is computed per test fold (4--5 samples, both classes present) and averaged over the three folds.
Pooled LOOCV AUROC is known to be biased, particularly under the null \cite{parker2007stratification,airola2011auc}; scheme (ii) is therefore the primary analysis and scheme (i) is reported for continuity with versions 1--3.

\paragraph{Variance.}
Under scheme (ii) we report the standard deviation of AUROC across training seeds within a partition (seed SD) and across partitions after averaging over seeds (split SD) \cite{bouthillier2021variance}.

\paragraph{Statistical comparisons.}
Differences in mean AUROC under scheme (ii) were tested with the corrected resampled $t$-test of Nadeau and Bengio \cite{nadeau2003inference,bouckaert2004evaluating} over the 30 partition-folds (seed-averaged).
Under scheme (i) we report bootstrap 95\% confidence intervals of the mean over seeds and make no formal comparisons, because of the bias of pooled LOOCV AUROC.
Differences in seed variance were summarised by the ratio of seed SDs within each partition, averaged over partitions, with a bootstrap 95\% confidence interval over partitions.
We avoid null-hypothesis tests based on five seeds, as used in versions 1--3, because they measure optimisation noise rather than generalisation \cite{bouthillier2021variance,demsar2006statistical}.

\paragraph{Label-permutation test.}
To test whether an arm learns anything at all, the full pipeline---selection and training---was re-run on 100 random permutations of the labels for one 3-fold partition and seed, and the observed AUROC was compared with the permutation distribution \cite{ojala2010permutation}; $p$-values are computed as $(b+1)/(m+1)$ \cite{phipson2010permutation}.

\paragraph{Sensitivity analyses.}
Ten-seed LOOCV runs were repeated (a) with 20 and 100 instead of 50 optimiser steps (warm-up scaled to 40\% of the steps), (b) with a LASER-VFL learning rate of $10^{-4}$, and (c) without the two possibly mislabelled samples ($N=11$).

\paragraph{Stability of selected features.}
The agreement of feature subsets selected by the same silo across training seeds (same fold) and across folds (same seed) was measured with the stability index of Nogueira et al.\ \cite{nogueira2018stability}, which is 0 for random selection and 1 for identical subsets.

\subsection*{Implementation and reproducibility}
Experiments ran on a workstation with two Intel Xeon Gold 6230 CPUs (80 threads) and 251 GB RAM, one CPU thread per fit, using Python 3.10, PyTorch 2.5.1 \cite{paszke2019pytorch}, scikit-learn 1.5.2 \cite{pedregosa2011sklearn}, NumPy \cite{harris2020numpy}, SciPy \cite{virtanen2020scipy}, pandas \cite{mckinney2010pandas} and Matplotlib \cite{hunter2007matplotlib}.
Every arm writes its per-sample test probabilities; all metrics are computed from those files.
The code for all experiments, which generates the fold assignments deterministically, is available at \url{https://github.com/samvictordr/REEF/tree/v4-benchmark/benchmark_v4}; the per-sample prediction files will be deposited in the same directory.

\section*{Results}

\subsection*{Reproduction of versions 1--3 and the source of their conclusions}
\label{sec:reproduction}
\begin{figure}[!ht]
\centering
\includegraphics[width=0.8\textwidth]{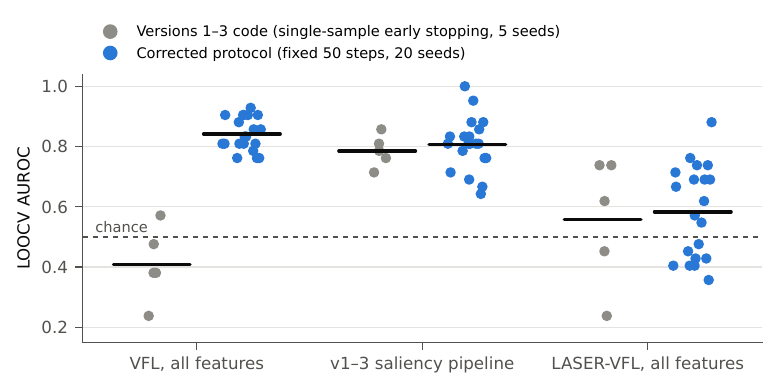}
\caption{{\bf Versions 1--3 versus the corrected protocol.}
Pooled-LOOCV AUROC per seed (points) and mean (bar).
Grey: original code, in which every model selected its checkpoint on a single validation sample (5 seeds).
Blue: fixed 50 optimiser steps, final model evaluated, identical data for all methods (20 seeds).
The VFL baseline's chance-level score came from evaluating it after one optimiser step.}
\label{fig:reproduction}
\end{figure}

Running the original benchmark code unchanged reproduced the published results closely (Fig~\ref{fig:reproduction}): pooled-LOOCV AUROC over the five original seeds was 0.786 $\pm$ 0.053 for the saliency pipeline (published 0.776 $\pm$ 0.039), 0.557 $\pm$ 0.213 for LASER-VFL (0.557 $\pm$ 0.191) and 0.410 $\pm$ 0.124 for the VFL baseline (0.500 $\pm$ 0.125).
The training logs show why.
Each training fold reserved one sample for validation, so validation AUROC was undefined and the code returned 0.5 in every fit.
The first round whose score exceeded the initial value of 0 was therefore kept as ``best'', and training stopped 15 rounds later.
All 65 VFL-baseline fits stopped at round 16 and were evaluated with their round-1 weights---a single optimiser step from random initialisation---and all 65 saliency-pipeline fits were evaluated after 21 steps (20 warm-up rounds plus one refinement round).
LASER-VFL monitored validation loss, which is defined for one sample; it stopped early in 19 of its 65 fits (between epochs 18 and 41) and was evaluated at the epoch with the lowest single-sample validation loss.
In addition, LASER-VFL received the unfiltered microbiome table (27,807 instead of 12,432 features) and mini-batches of 8 that dropped 3 of the 11 training samples in every epoch.

When every model is instead trained for the same 50 steps on the same data and evaluated at the end of training, the VFL baseline is no longer at chance (0.842 $\pm$ 0.055 over 20 seeds) and its mean exceeds that of the saliency pipeline (0.807 $\pm$ 0.088; Fig~\ref{fig:reproduction}).
The central claims of versions 1--3---that standard VFL fails at chance level, that the saliency pipeline reduces variance three- to five-fold, and that its biological priors are the source of that stability---were artefacts of this evaluation and are withdrawn.

\subsection*{Benchmark: no selection strategy or prior separates from plain VFL}
\begin{figure}[!ht]
\centering
\includegraphics[width=\textwidth]{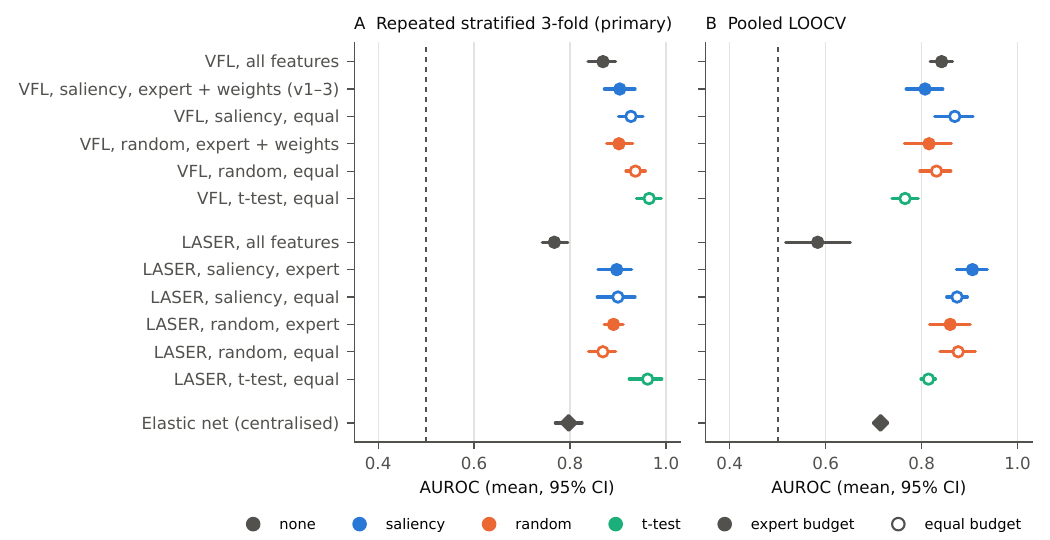}
\caption{{\bf Benchmark results.}
Mean AUROC and 95\% bootstrap confidence interval for every arm under (A) repeated stratified 3-fold cross-validation (10 partitions $\times$ 10 seeds; interval over partitions) and (B) pooled LOOCV (20 seeds; interval over seeds).
Colour: feature-selection strategy; filled markers: expert budget; open markers: equal budget; diamond: centralised elastic net. Dashed line: chance.}
\label{fig:benchmark}
\end{figure}

Table~\ref{tab:main} and Fig~\ref{fig:benchmark} summarise all arms.
Under the primary scheme (repeated stratified 3-fold), every VFL arm reached a mean AUROC between 0.87 and 0.97.
The saliency pipeline of versions 1--3 (0.904; 95\% CI over partitions 0.872--0.935) was indistinguishable from VFL on all features (0.869, 0.838--0.895; corrected $p = 0.73$), from saliency selection with equal budgets (0.927; $p = 0.56$), from random selection with the same budgets and embedding weights (0.902; $p = 0.97$) and from random selection with equal budgets (0.936; $p = 0.58$).
The label-aware $t$-test filter had the highest mean (0.965) but did not differ significantly either ($p = 0.36$).
Under pooled LOOCV the ordering changed---VFL on all features (0.842) ranked above the saliency pipeline (0.807), and the $t$-test filter fell to 0.765---which shows how little either scheme can resolve with 13 samples \cite{varoquaux2018cv,airola2011auc}.
The corrected resampled $t$-test is deliberately conservative \cite{nadeau2003inference,bouckaert2004evaluating}; the defensible summary is that none of the VFL variants can be distinguished from one another, including the variant with no feature selection at all.

With the learning rate used in versions 1--3 ($10^{-3}$), LASER-VFL trained on all 90,579 features was the weakest neural arm under both schemes (3-fold 0.767; LOOCV 0.583) and the least stable across seeds.
Given a 1,300-feature subset it performed on par with the VFL arms (3-fold 0.868--0.962, LOOCV 0.814--0.906), and random subsets were about as effective as saliency subsets.
This apparent benefit of masking turned out to be a learning-rate effect (see \nameref{sec:sensitivity}).
The centralised elastic net reached 0.797 (3-fold) and 0.714 (LOOCV), at or below the neural VFL arms.

\begin{table}[!ht]
\begin{adjustwidth}{-2.25in}{0in}
\centering
\caption{{\bf AUROC of every arm.} 3-fold: repeated stratified 3-fold cross-validation, 10 partitions $\times$ 10 seeds (mean, 95\% bootstrap CI over partitions, SD across seeds within partition, SD across partitions). LOOCV: pooled leave-one-out AUROC over 20 seeds (mean, 95\% bootstrap CI and SD over seeds). All neural arms trained for 50 steps. $^\dagger$Configuration of versions 1--3. $^\ddagger$Requires labels at each silo.}
\label{tab:main}
\footnotesize
\begin{tabular}{lll cccc cc}
\toprule
 & & & \multicolumn{4}{c}{3-fold (primary)} & \multicolumn{2}{c}{LOOCV} \\
\cmidrule(lr){4-7}\cmidrule(lr){8-9}
Backbone & Selection & Budget & Mean & 95\% CI & Seed SD & Split SD & Mean [95\% CI] & SD \\
\midrule
VFL & none & -- & 0.869 & 0.838--0.895 & 0.050 & 0.048 & 0.842 [0.818, 0.864] & 0.055 \\
VFL & saliency & expert + weights$^\dagger$ & 0.904 & 0.872--0.935 & 0.070 & 0.054 & 0.807 [0.768, 0.844] & 0.088 \\
VFL & saliency & equal & 0.927 & 0.901--0.952 & 0.071 & 0.043 & 0.869 [0.827, 0.907] & 0.094 \\
VFL & random & expert + weights & 0.902 & 0.876--0.931 & 0.080 & 0.047 & 0.815 [0.764, 0.862] & 0.114 \\
VFL & random & equal & 0.936 & 0.917--0.957 & 0.071 & 0.034 & 0.831 [0.796, 0.861] & 0.074 \\
VFL & $t$-test$^\ddagger$ & equal & 0.965 & 0.938--0.990 & 0.021 & 0.044 & 0.765 [0.738, 0.793] & 0.063 \\
\midrule
LASER-VFL & none & -- & 0.767 & 0.742--0.795 & 0.112 & 0.046 & 0.583 [0.517, 0.651] & 0.154 \\
LASER-VFL & saliency & expert & 0.897 & 0.858--0.928 & 0.090 & 0.060 & 0.906 [0.873, 0.937] & 0.075 \\
LASER-VFL & saliency & equal & 0.900 & 0.857--0.935 & 0.077 & 0.066 & 0.874 [0.852, 0.895] & 0.051 \\
LASER-VFL & random & expert & 0.891 & 0.871--0.911 & 0.096 & 0.034 & 0.860 [0.817, 0.901] & 0.100 \\
LASER-VFL & random & equal & 0.868 & 0.839--0.895 & 0.121 & 0.048 & 0.876 [0.838, 0.912] & 0.087 \\
LASER-VFL & $t$-test$^\ddagger$ & equal & 0.962 & 0.923--0.991 & 0.010 & 0.060 & 0.814 [0.799, 0.829] & 0.035 \\
\midrule
Elastic net (central) & -- & -- & 0.797 & 0.769--0.825 & 0 & 0.047 & 0.714 [0.714, 0.714] & 0.000 \\
\bottomrule
\end{tabular}
\end{adjustwidth}
\end{table}

\subsection*{Seed variance and split variance}
\begin{figure}[!ht]
\centering
\includegraphics[width=0.85\textwidth]{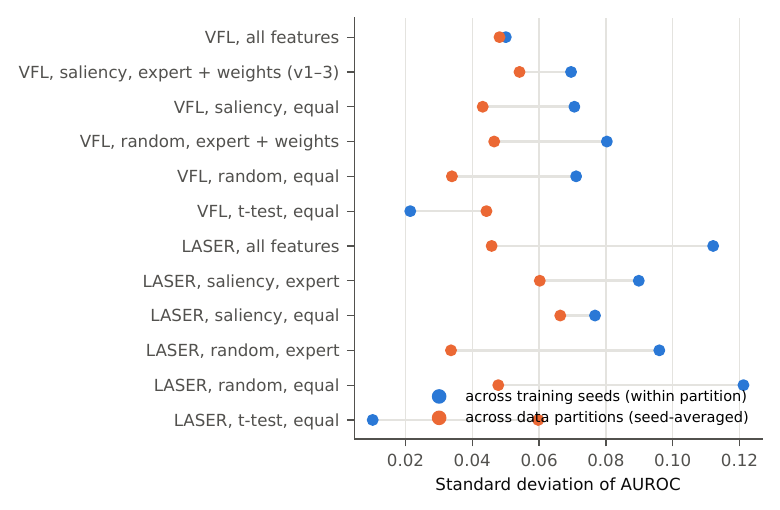}
\caption{{\bf Seed and split variability (repeated stratified 3-fold).}
Blue: standard deviation of AUROC across 10 training seeds within a data partition, averaged over partitions. Orange: standard deviation across 10 partitions of the seed-averaged AUROC.}
\label{fig:variance}
\end{figure}

Fig~\ref{fig:variance} separates the two sources of variability under the 3-fold scheme.
For the VFL arms the standard deviation across training seeds (0.05--0.08) was of the same order as the standard deviation across data partitions (0.03--0.05).
The saliency pipeline's seed SD (0.070) was not lower than that of VFL on all features (0.050; mean per-partition ratio of seed SDs, VFL/saliency, 0.82, bootstrap 95\% CI over partitions 0.61--1.04) or of random selection (0.071--0.080).
Two groups stood out.
LASER-VFL at learning rate $10^{-3}$ tended to have larger seed variability, with or without masks (seed SD 0.077--0.121); on all features its mean per-partition ratio to the saliency pipeline was 1.95 (95\% CI 1.32--2.75).
The two $t$-test arms, whose feature subsets do not depend on the seed, had the smallest (0.021 and 0.010).
Seed variability therefore tracks how much of the pipeline depends on the random seed, not the presence of biological priors.

\subsection*{Gradient saliency selects by initialisation, not by signal}
If gradient saliency identified stress-related features, the same silo should select overlapping features when only the random seed changes.
It did not (Table~\ref{tab:stability}): across seeds within a fold, the stability index of saliency subsets was 0.00--0.06, indistinguishable from random selection, whereas the $t$-test selected identical subsets (1.00).
Across folds within a seed---where the initial weights are shared but the training samples differ---saliency subsets overlapped more (0.13--0.30).
The saliency ranking is thus driven mainly by the network's random initial weights, consistent with sanity checks showing that gradient attributions can reflect architecture and initialisation rather than data \cite{adebayo2018sanity}.
With a learning rate of $5\times10^{-5}$ and 20 warm-up steps, the encoders move little from initialisation before the scores are computed.
Consequently, feature lists produced by the saliency pipeline cannot be interpreted as biomarkers, and the omic-level ``saliency hierarchy'' reported in versions 1--3 (proteomics dominating transcriptomics) is not evidence about biology.

\begin{table}[!ht]
\begin{adjustwidth}{-2.25in}{0in}
\centering
\caption{{\bf Stability of selected feature subsets (Nogueira index; LOOCV, 20 seeds $\times$ 13 folds).} Across seeds: same fold, different training seeds. Across folds: same seed, different training folds. 0 corresponds to random selection, 1 to identical subsets.}
\label{tab:stability}
\footnotesize
\begin{tabular}{l cc cc cc}
\toprule
 & \multicolumn{2}{c}{Saliency, expert$^\dagger$} & \multicolumn{2}{c}{Saliency, equal} & \multicolumn{2}{c}{$t$-test, equal} \\
\cmidrule(lr){2-3}\cmidrule(lr){4-5}\cmidrule(lr){6-7}
Silo & seeds & folds & seeds & folds & seeds & folds \\
\midrule
Transcriptomics & 0.05 & 0.13 & 0.06 & 0.14 & 1.00 & 0.56 \\
Proteomics & 0.00 & 0.28 & 0.00 & 0.30 & 1.00 & 0.73 \\
Metabolomics & 0.00 & 0.23 & 0.00 & 0.20 & 1.00 & 0.52 \\
Microbiome & 0.02 & 0.14 & 0.02 & 0.14 & 1.00 & 0.70 \\
\bottomrule
\end{tabular}
\end{adjustwidth}
\end{table}

\subsection*{Permutation test}
\begin{figure}[!ht]
\centering
\includegraphics[width=\textwidth]{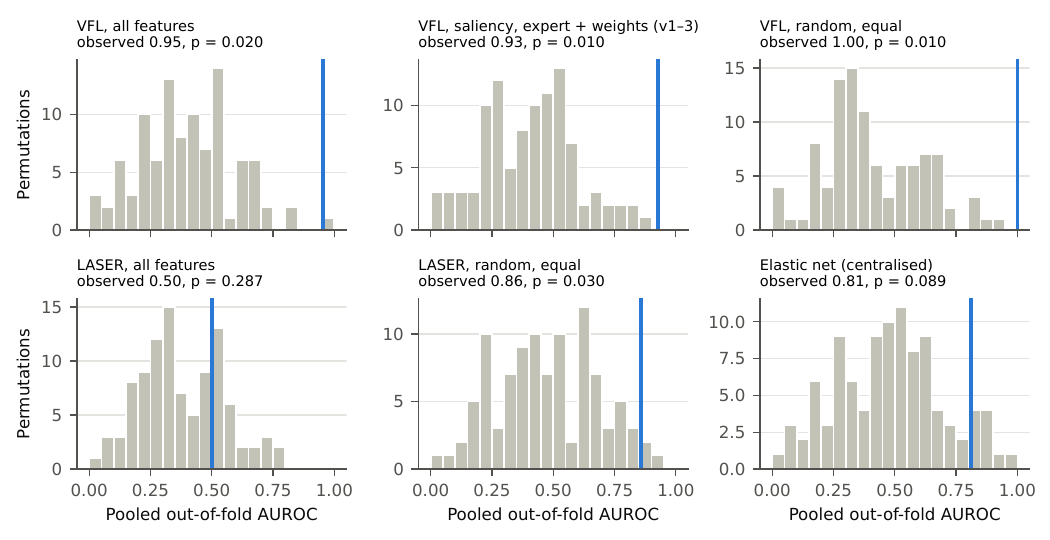}
\caption{{\bf Label-permutation test.}
Grey: pooled out-of-fold AUROC of the complete pipeline (selection and training) on 100 label permutations, for one stratified 3-fold partition and seed. Blue line: AUROC with the true labels.}
\label{fig:permutation}
\end{figure}

For one 3-fold partition and seed, the full pipeline was re-run on 100 label permutations (Fig~\ref{fig:permutation}).
Because a permuted test fold can contain a single class, the statistic is the AUROC of the 13 pooled out-of-fold predictions.
VFL on all features (observed 0.95, $p = 0.020$), the saliency pipeline (0.93, $p = 0.010$), random selection with equal budgets (1.00, $p = 0.010$) and LASER-VFL with random masks (0.86, $p = 0.030$) all exceeded their permutation distributions; $p = 0.010$ is the smallest value attainable with 100 permutations \cite{phipson2010permutation}.
LASER-VFL on all features at learning rate $10^{-3}$ did not (0.50, $p = 0.29$), and the elastic net was borderline (0.81, $p = 0.089$).
The permutation distributions were centred near 0.40 rather than 0.50, the downward bias expected for pooled cross-validated AUROC under the null \cite{parker2007stratification,airola2011auc}; the sub-chance negative controls reported in versions 1--3 are an instance of this bias.

\subsection*{Sensitivity analyses}
\label{sec:sensitivity}
\begin{figure}[!ht]
\centering
\includegraphics[width=\textwidth]{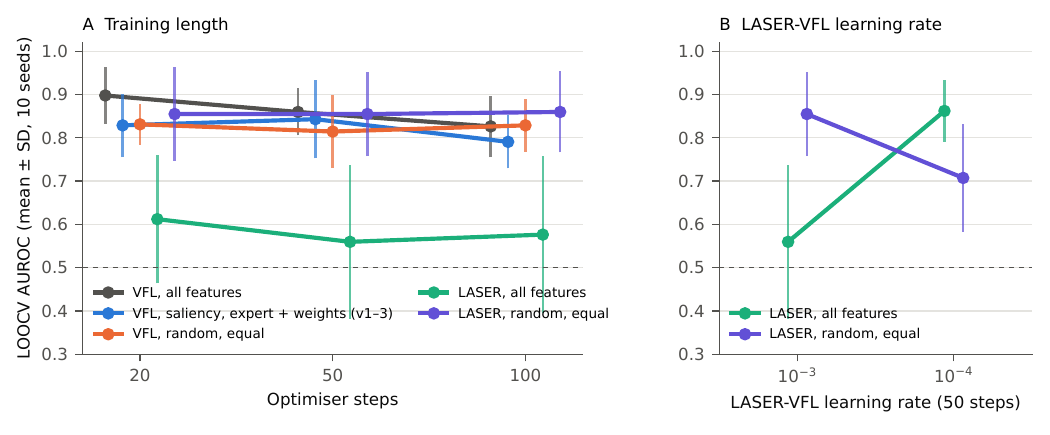}
\caption{{\bf Sensitivity analyses (pooled LOOCV, 10 seeds, mean $\pm$ SD).}
(A) Number of optimiser steps (for the saliency pipeline, warm-up is 40\% of the steps). (B) LASER-VFL learning rate at 50 steps.}
\label{fig:sensitivity}
\end{figure}

Training length did not change the picture (Fig~\ref{fig:sensitivity}A): over 20, 50 and 100 steps, VFL on all features (0.90, 0.86, 0.83), the saliency pipeline (0.83, 0.84, 0.79) and random selection (0.83, 0.81, 0.83) stayed within a few hundredths of one another, and LASER-VFL at learning rate $10^{-3}$ on all features stayed low (0.61, 0.56, 0.58).

LASER-VFL's learning rate did (Fig~\ref{fig:sensitivity}B).
At $10^{-4}$, LASER-VFL on all features reached 0.862 $\pm$ 0.072, comparable to every VFL arm, whereas LASER-VFL with random masks dropped from 0.855 to 0.707.
The failure of LASER-VFL on all features reported in versions 1--3 and in Table~\ref{tab:main} is therefore a mismatch between learning rate and input dimensionality---a plausible explanation is that, with Adam-type optimisers, each step changes every first-layer weight by roughly the learning rate, so the change in pre-activations grows with the number of inputs---and not evidence that LASER-VFL, or VFL in general, needs feature selection in this regime.

Excluding the two possibly mislabelled samples ($N = 11$; 5 heat-stressed) left the conclusions unchanged: pooled-LOOCV AUROC over 10 seeds was 0.953 $\pm$ 0.067 for VFL on all features, 0.913 $\pm$ 0.077 for random selection, 0.823 $\pm$ 0.186 for the saliency pipeline, 0.860 $\pm$ 0.102 for LASER-VFL with random masks, 0.617 $\pm$ 0.135 for LASER-VFL on all features at learning rate $10^{-3}$, and 0.833 for the elastic net.

\section*{Discussion}

\subsection*{What makes VFL work with 13 samples}
On this dataset, a plain split-network VFL model trained on all 90,579 features for a fixed number of steps already separates heat-stressed from non-stressed samples well above chance, as confirmed by an end-to-end permutation test.
Neither component of the saliency pipeline of versions 1--3---gradient-saliency selection and hand-set priors over omics layers---produced a difference that the data could resolve, and random feature subsets of the same size did as well.
LASER-VFL appeared to need feature masking, but only at the learning rate used in versions 1--3; at a ten-fold lower learning rate it matched the other methods on all features.
In other words, every intervention that looked important in versions 1--3 turned out to be either an evaluation artefact or a hyperparameter artefact.
What did matter was giving each model a training configuration appropriate to its input dimension and evaluating all of them the same way.

\subsection*{Lessons for evaluating VFL with very few samples}
The failure of versions 1--3 is an instance of well-documented pitfalls rather than a new one.
Checkpoint selection on a validation set that is too small to be informative silently turns ``early stopping'' into ``evaluate after an arbitrary number of steps'' \cite{prechelt1998early,cawley2010overfitting}; in our case it differed by method, which created a large spurious effect.
Baselines that are trained or evaluated with less care than the proposed method are a recurrent source of overstated gains \cite{melis2018state,lipton2019troubling,kapoor2023leakage}.
Five seeds and paired $t$-tests over seeds quantify optimisation noise, not generalisation \cite{bouthillier2021variance,henderson2018deep}, and pooled LOOCV AUROC is biased and unstable \cite{parker2007stratification,airola2011auc}; the sub-chance permutation results of versions 1--3 (AUROC 0.36 and 0.24), which were interpreted as ``anti-learning'' caused by biological priors, are consistent with this known bias.
Finally, demonstrating that a feature-selection method outperforms random selection is a necessary control: random gene signatures are often as predictive as published ones \cite{venet2011random}.
We recommend that VFL studies with $N < 100$ (i) fix training length or tune it in an inner loop, never on a handful of validation samples; (ii) give every method the same data, step budget and tuning budget; (iii) report seed and split variability separately; (iv) include random-selection and centralised baselines; and (v) run end-to-end permutation tests.
These recommendations follow general guidance for neutral benchmarking and reproducible machine learning \cite{boulesteix2013plea,weber2019essential,pineau2021improving,gundersen2018state}.

\subsection*{Limitations}
This is a single dataset with 13 samples from a single colony, so every estimate is uncertain and none of our conclusions should be extrapolated to other coral species, stressors or omics designs without replication.
The label combines time points; T1 samples were taken shortly after the temperature ramp ended, and field samples were treated as non-stressed, following their reported similarity to ambient samples \cite{williams2023peeling}.
Two samples may be mislabelled \cite{williams2023peeling}; excluding them did not change any conclusion, and plain VFL remained the best or joint-best arm.
No hyperparameters were tuned systematically; the LASER-VFL learning-rate result shows that a single untuned setting can change a method's ranking, so a nested tuning loop with equal budgets per method \cite{bergstra2012random,cawley2010overfitting} is the natural next step; LASER-VFL in particular was used outside its design setting of missing feature blocks \cite{valdeira2025laser}.
The VFL simulation ran in a single process; we did not evaluate communication cost, privacy leakage through embeddings or gradients \cite{zhu2019dlg,luo2021feature,fu2022label}, or defences such as differential privacy \cite{abadi2016dpsgd,dwork2014algorithmic}.
The label-aware $t$-test selector requires labels at every silo and is therefore not a valid federated method without additional protocols \cite{castiglia2023lessvfl,li2023fedsdgfs}.
Benchmarks of VFL across many datasets, such as VertiBench \cite{wu2024vertibench}, and of multi-omics integration methods \cite{cantini2021benchmarking,herrmann2021benchmark,leng2022benchmark} are needed to establish which of our observations generalise.

\section*{Conclusion}
In a controlled benchmark on a 13-sample coral thermal-stress multi-omics dataset, vertical federated learning worked well above chance without domain priors or informed feature selection.
Gradient-saliency selection behaved like random selection, hand-set priors over omics layers had no detectable effect, and the apparent failure of LASER-VFL on all features disappeared with a smaller learning rate.
The conclusions of the earlier versions of this work were caused by checkpoint selection on single-sample validation sets and are withdrawn.
For small-sample federated multi-omics, careful evaluation matters more than method design: the corrected baseline was as good as the method proposed to replace it.

\section*{Data and code availability}
The dataset is available from Zenodo \cite{stephens2022zenodo}. Code for all experiments, which generates the fold assignments deterministically, is available at \url{https://github.com/samvictordr/REEF/tree/v4-benchmark/benchmark_v4}. The per-sample prediction files will be deposited in the same directory.

\section*{Supporting information}
\paragraph*{S1 Text.}
\label{S1_Text}
{\bf LASER-VFL adaptation details.} Modifications made to the LASER-VFL reference implementation for numerical stability, and files restored from the upstream repository.

\section*{Acknowledgments}
The author thanks Williams, Stephens and colleagues for making the \textit{Montipora capitata} multi-omics data publicly available \cite{williams2021multiomic,stephens2022zenodo}, the authors of LASER-VFL for releasing their code, and Swarn S.~W.\ for proofreading an earlier version of the manuscript.
The author used AI-assisted tools (Claude, Anthropic) to audit the code of versions 1--3, run and analyse the corrected experiments, verify references, and draft and edit this version; the author reviewed and takes responsibility for all content.
The author is a collaborator with Rikoukei Labs FZ-LLC, a startup working on multi-omics platforms. This research received no institutional or corporate funding, data or computational resources. The author declares no competing interests.

\bibliography{references}

\end{document}